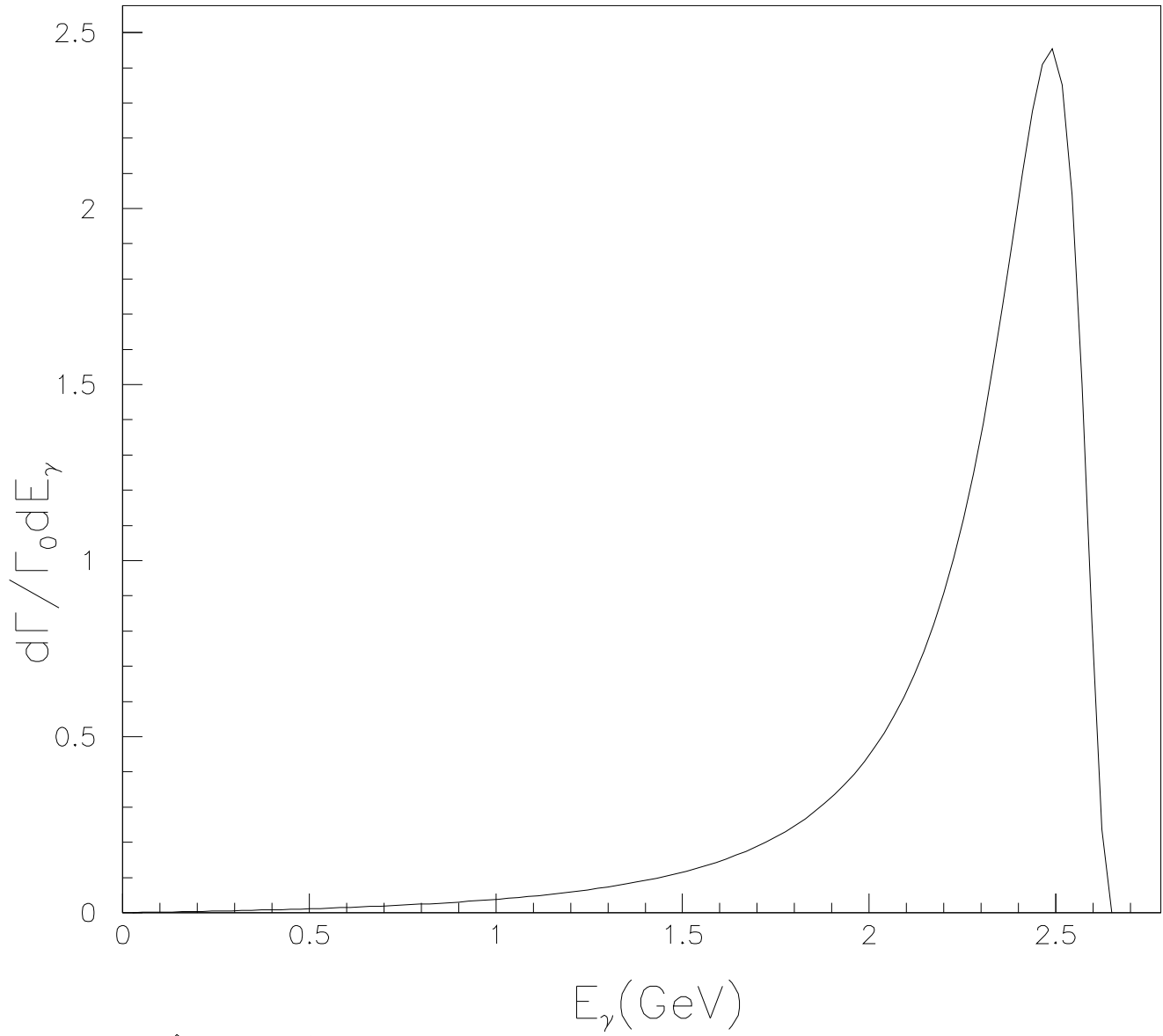

Fig. 1

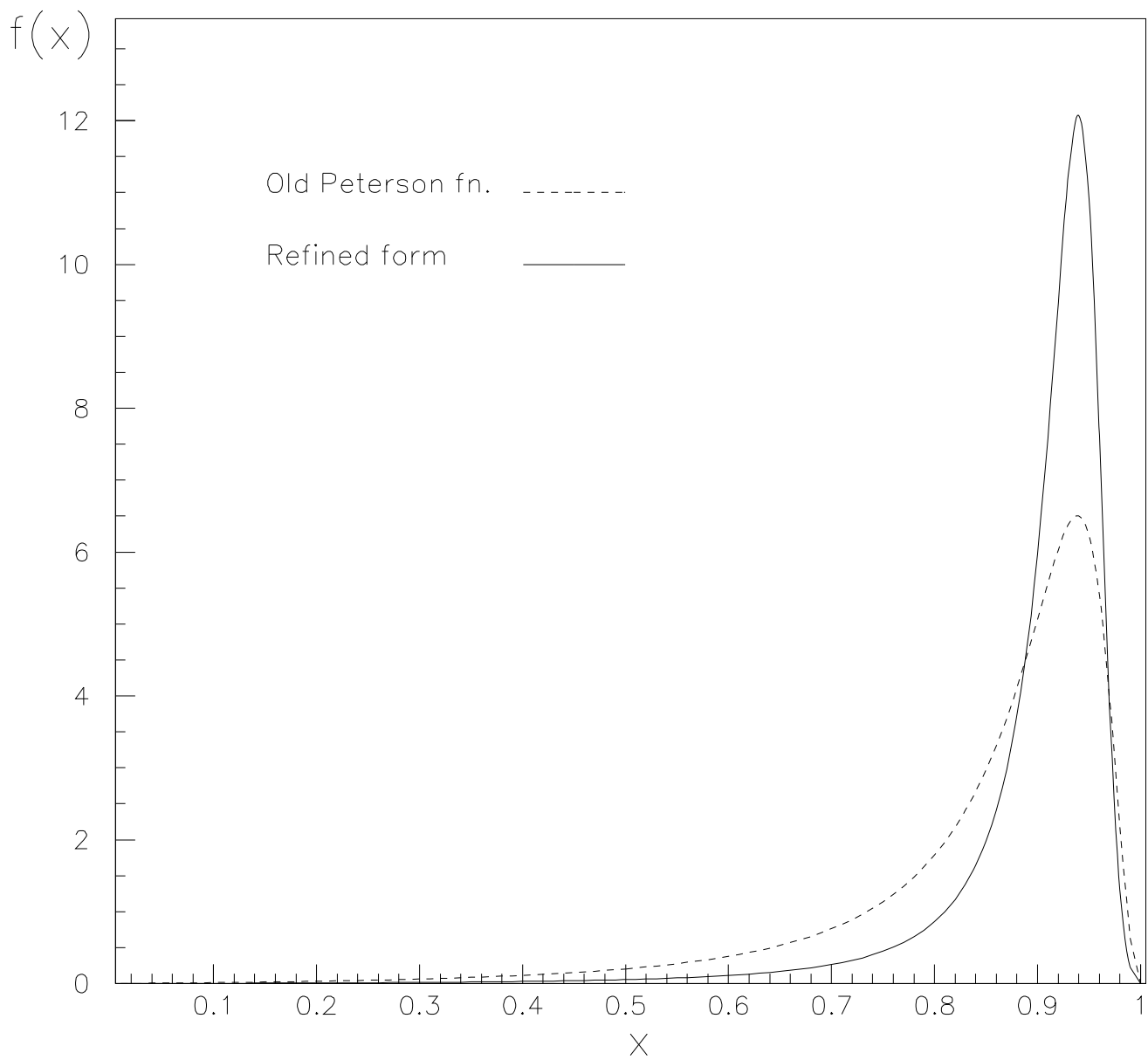

Fig. 2

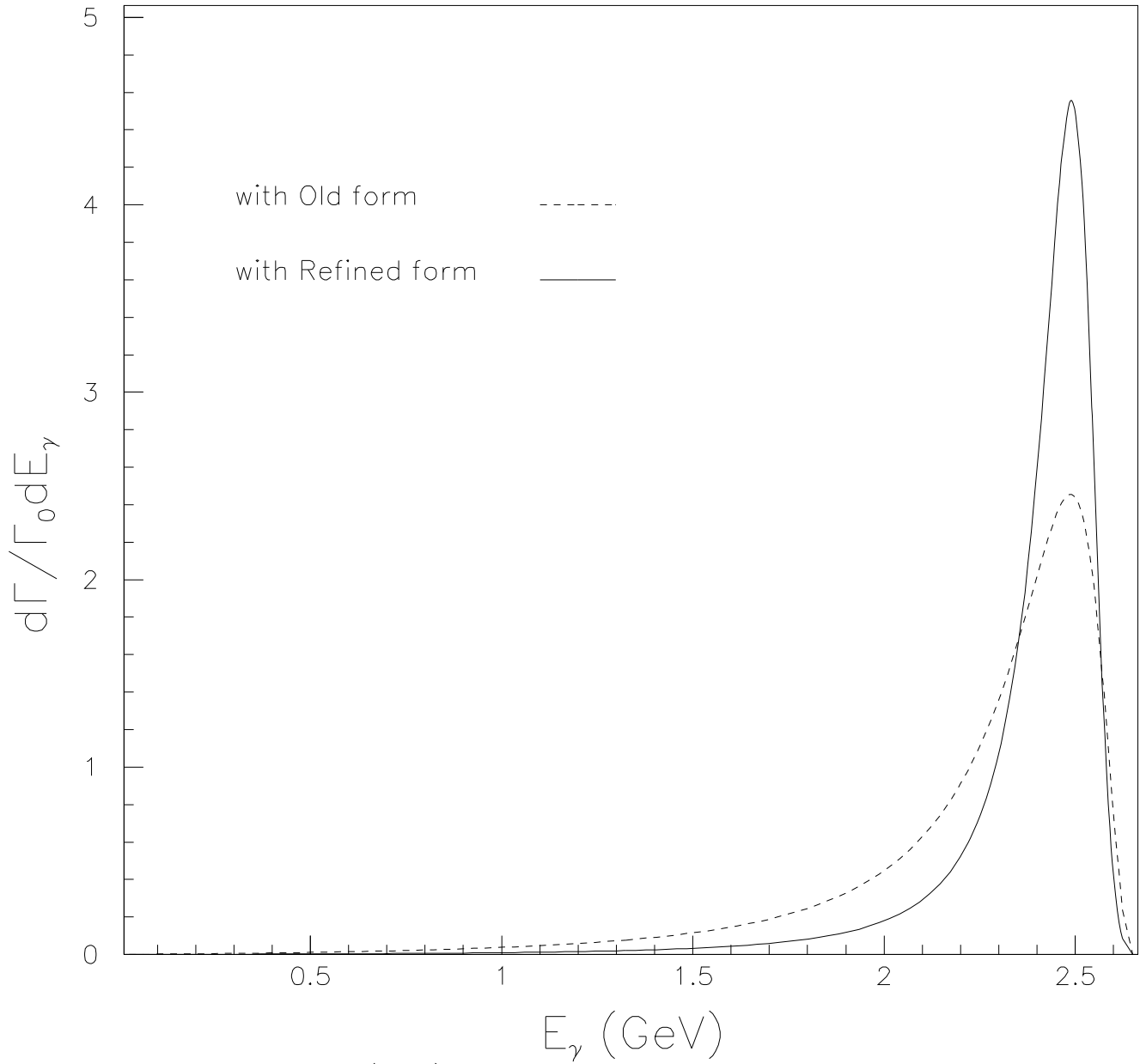

Fig. 3 (a)

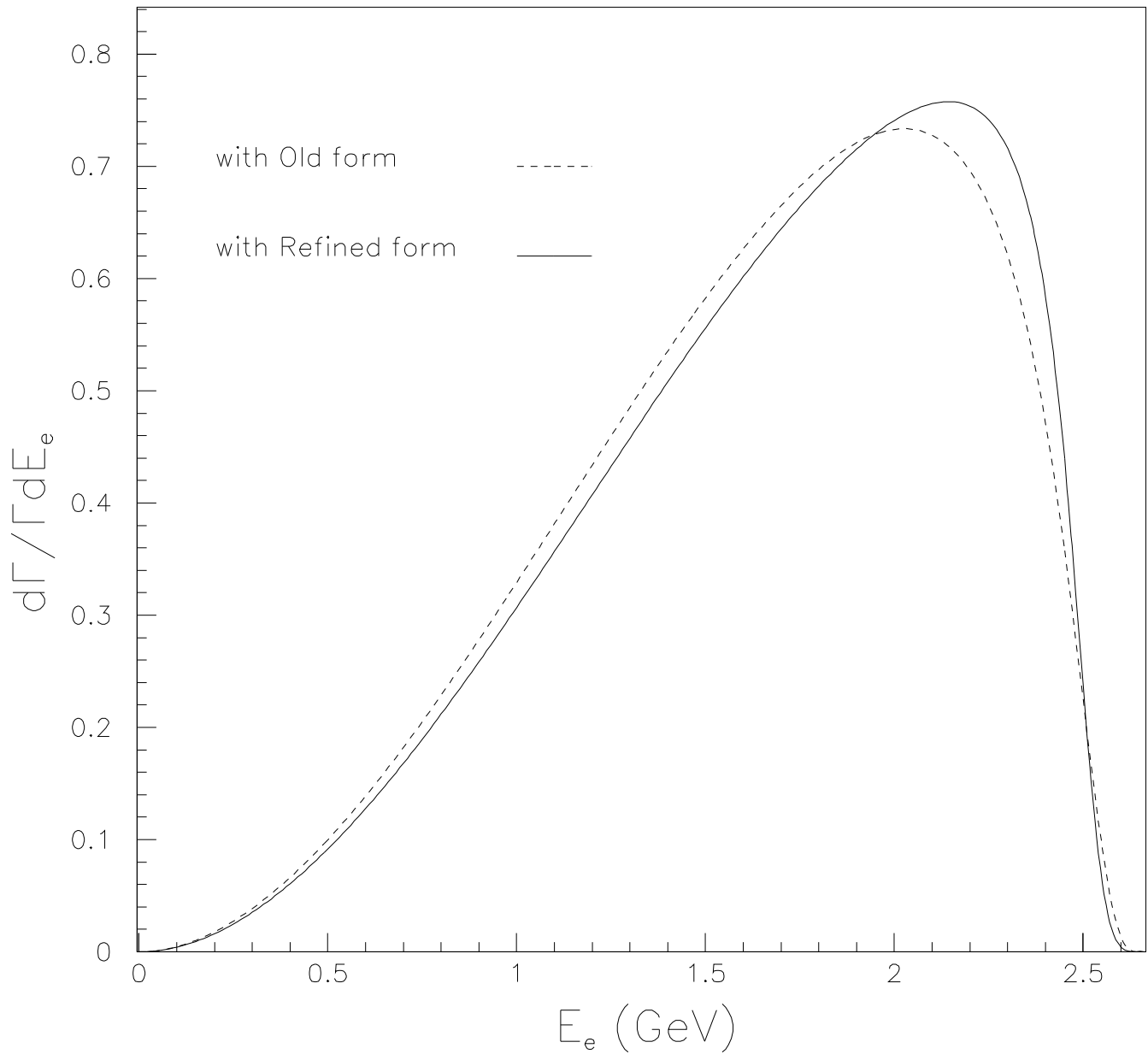

Fig. 3 (b)

Figure 1: The photon energy spectrum of $B \to \gamma X_s$ process calculated by the parton model.

Figure 2: Peterson's fragmentation function and our refined form for describing decay processes. The solid line denotes our refined function and the dashed line Peterson's original function.

Figure 3: (a) The photon energy spectrum of $B \to \gamma X_s$ process with Peterson's function (dashed line) and with our refined function (solid line). (b) The electron energy spectrum of $B \to e\bar{\nu}X_u$ process with Peterson's function (dashed line) and with our refined function (solid line).

Table 1: The second moments of line shape $a_2$, average kinetic energies of $b$–quark, effective masses of light degrees of freedom and $b$–quark masses calculated with our refined function varying the parameter $\alpha_Q$.

| $\alpha_Q$ | $a_2$ | $\mu_\pi^2$ | $\bar{\Lambda}$ | $m_b$ |
|---|---|---|---|---|
| 0.02 | 0.762 | 1.290 | 0.751 | 4.549 |
| 0.04 | 0.759 | 1.094 | 0.693 | 4.607 |
| 0.06 | 0.742 | 0.880 | 0.629 | 4.671 |
| 0.08 | 0.693 | 0.645 | 0.557 | 4.743 |
| 0.085 | 0.673 | 0.583 | 0.537 | 4.763 |
| 0.1 | 0.575 | 0.388 | 0.475 | 4.825 |

# The Parton Model Approach to Inclusive $B$ meson Decays

Kang Young Lee* and Jae Kwan Kim

*Department of Physics, Korea Advanced Institute of Science and Technology*

*Taejon 305–701, KOREA*

(February 18, 1996)

## Abstract

We test the parton model approach to inclusive $B \to \gamma X_s$ and $B \to e \bar{\nu} X_u$ processes by calculating a few moments of the distribution function. The $1/m_Q$ expansion of $b$-quark distribution function is discussed. We show that Peterson's fragmentation function is useful as the distribution function of $b$-quark but requires some improvements. By extension of Peterson's arguments, we obtain the refined form for the distribution function and with the refined function, we obtain the consistent results with the HQET.

Typeset using REVTEX

*kylee@chep5.kaist.ac.kr

# I. INTRODUCTION

The parton model approach has been established as a phenomenological model to describe inclusive semileptonic $B$ decays [1,2] and found to give good agreement with experiments for electron energy spectrum at all energies. The parton model is motivated by the deep inelastic scattering process (DIS) and it is well known that inclusive semileptonic decays of heavy flavours are intimately related to DIS via channel crossing. Jin and Paschos gave a theoretical foundation on this model recently [3]. The light–cone dominance of $B$ decay processes allows one to use the methods of DIS and express the commutator of two currents as bilocal operator of quark fields. Its Fourier transform is related to the distribution function of $b$–quark inside $B$ meson. Nonperturbative effects of QCD are included in the matrix elements of the bilocal operator and also in the distribution function.

Over the past few years, there has been much progress in the study of inclusive semileptonic decays of $B$ mesons with the help of heavy quark effective theory (HQET) [4–6]. One can get a QCD-based expansion in powers of $1/m_Q$ by performing an operater product expansion on the hadronic tensor. With this technique, the systematic treatments of the inclusive spectra is achieved in principle. Therefore it is important to reanalyze phenomenological models in the framework of the HQET.

The ACCMM model [7], which is the most popular model for the inclusive semileptonic decays of heavy mesons, has been reanalyzed in the framework of the HQET recently by several authors [8,9]. They showed that ACCMM function for certain values of parameter can be viewed as good approximation for leading twist nonperturbative contribution to differential distribution curves.

In this paper we analyze the parton model with the help of the HQET. Since the parton model is constructed in the analogy of DIS, we can study the model in the context of the HQET more easily than the ACCMM model. With properly defined $m_b$, the parton model is represented by $1/m_b$ expansion, and the distribution function of $b$–quark inside a $B$ meson used in this model is directly related to the genuine QCD distribution function which is not obtained systematically. Firstly we use Peterson's fragmentation function as the distribution function of $b$-quark following the original works of ref. [1,2]. QCD cannot determine the genuine distribution function completely, but only parametrize its moments. Thus we represent the function as infinite sum of singular functions in power of $1/m_Q$ with the moments. By calculating the moments of the QCD distribution function, we test the validity of the model. We find some discrepancies in the values of the parameters $\mu_\pi^2$ and $\bar\Lambda$ obtained by the parton model from those of the HQET while the parton model gives well agreed decay spectra with those of the HQET. Hence an improvement of the functional form is required. We refine the functional form by extending Peterson's approach to the decay process and obtain the consistent result with the HQET.

This paper is organized as follows: in section 2 we formulate $1/m_Q$ expansion of the distribution function. In section 3, we derive the differential decay rate of inclusive $B \to \gamma X_s$ decay in the parton model and show that the $b$–quark distribution function of the parton model is corresponding to the genuine distribution function of QCD. The inclusive semileptonic $b \to u$ decays are also studied. We test the distribution function by calculating a few moments in section 4. In section 5, we improve the functional form of Peterson's function. Our conclusions are summarized in section 6.

## II. HEAVY QUARK EXPANSION OF THE PARTON MODEL

The parton model approach suggested in ref. [1,2] has been established on the bases of following two postulations:

$i$) This model pictures the mesonic decay as the decay of the partons with the incoherent assumption in analogy to deep inelastic scattering process. The probability of finding a $b$-quark in a $B$ meson carrying a fraction $x$ of the meson momentum in the infinite momentum frame is given by the distribution function $f(x)$. The decay probability is determined by the momentum of $b$–quark inside the $B$ meson and has the value of $f(x)dx$. Then we write the Lorentz invariant decay width as follows:

$$E_B \, d\Gamma(B \to X_q e\nu) = \int dx \, f(x) \, E_b \, d\Gamma(b \to q e\nu) \quad, \tag{1}$$

with the relation $p_b = x p_B$.

*ii*) Peterson's fragmentation function [10] is used as the distribution function $f(x)$ of $b$–quark because we expect that the functional forms of the distribution function and fragmentation function are similar for heavy flavours. This is the fragmentation function usually used in Lund Monte Carlo programs:

$$f_Q(x) = \frac{N_Q}{x\left(1 - \frac{1}{x} - \frac{\epsilon_Q}{1-x}\right)^2} \tag{2}$$

where $\epsilon_Q$ is a parameter dependent upon the heavy flavor and $N_Q$ is the corresponding normalization factor.

The first postulation has been studied by several authors [11,12]. The universal distribution function of the QCD is interpreted as the structure function of the $b$–quark in the parton model, which determines the distribution of the light–cone projection of the $b$–quark momentum inside the $B$ meson. In this and the next sections, we present the heavy quark expansion of the model in the viewpoint of the parton model.

Being on a ground of an phenomenological model, the $b$–quark mass $m_b$ is not an ingredient of the model and should be defined in the framework of the model in order to express the phenomenological model as $1/m_Q$ expansion. We define the $b$–quark mass as:

$$m_b \equiv \langle x \rangle m_B \quad, \tag{3}$$

following ref. [12]. Since we have the measured value of $m_B$ and well–defined function $f(x)$, this relation explicitly defines the mass of the $b$–quark. The nonperturbative informations of the parton model are encoded in the distribution function $f(x)$. If we want to express the parton model with the heavy quark expansion, it is performed by expanding the distribution function in $1/m_Q$. We write the expansion of the distribution function $f(x)$ as

$$f(x) \, dx = N \left( \delta(x_B) + \frac{1}{m_b} f_1(x_B) + \frac{1}{m_b^2} f_2(x_B) + ... \right) dx_B \tag{4}$$

where $N$ is a normalization constant, which is expressed by the variable

$$x_B = \frac{m_B}{\bar{\Lambda}}(x - x_0) \tag{5}$$

chosen by Bigi et al. [9,11]. Here $\bar{\Lambda}$ is the effective mass of light degrees of freedom inside $B$ meson and $x_0 = 1 - \bar{\Lambda}/m_B$. The leading term is the Dirac delta function since we demand that the leading term of $1/m_b$ expansion gives the free quark decay when the normalization constant is also written in the form of $1/m_b$ expansion

$$N = \left(1 + \frac{1}{m_b}\int dx_B\ f_1(x_B) + \frac{1}{m_b^2}\int dx_B\ f_2(x_B) + ...\right)^{-1}.$$

Our definition of $m_b$ results in the well–known relation of meson and quark masses in the parton model. From the definition of the $b$–quark mass, we obtain

$$\begin{aligned}m_B = m_b + \bar{\Lambda} &- \frac{\bar{\Lambda}}{m_b}\int dx_B\ x_B f_1(x_B) \\ &- \frac{\bar{\Lambda}}{m_b^2}\left(\int dx_B\ x_B f_2(x_B) - \int dx_B\ x_B f_1(x_B)\int dx_B\ f_1(x_B)\right) \\ &+ \mathcal{O}\left(\frac{1}{m_b^3}\right).\end{aligned} \tag{6}$$

Provided we let

$$\bar{\Lambda}\int dx_B\ x_B\ f_1(x_B) = \Delta m_B^2, \tag{7}$$

we obtain the mass relation

$$m_B = m_b + \bar{\Lambda} - \frac{\Delta m_B^2}{m_b} + \mathcal{O}(\frac{1}{m_b^2}). \tag{8}$$

## III. INCLUSIVE SPECTRA AND THE DISTRIBUTION FUNCTION

In this section, we explicitly relate the universal distribution function of the QCD to the parton structure function in order to help us to calculate the moments of the distribution function of $b$–quark in the framework of the parton model.

When we do not consider the perturbative QCD corrections, we can investigate the decay spectrum in cleaner way in the inclusive radiative decay process $B \to \gamma X_s$ than in the inclusive semileptonic decays. This process is not studied with the parton model framework yet. The relevant effective hamiltonian is given by

$$\mathcal{H}_{eff} = -\frac{4G_F}{\sqrt{2}} V_{tb} V_{ts}^* c_7(m_b) \frac{e}{16\pi^2} \bar{s} \sigma^{\mu\nu} (m_b P_R + m_s P_L) b F_{\mu\nu} \tag{9}$$

in leading log approximation. Hereafter we neglect the $s$–quark mass. Since the partonic subprocess is 2–body decay, the photon energy spectrum is represented by a monochromatic line:

$$\frac{d\Gamma^{(0)}}{dE_\gamma} = \Gamma_0 \delta(E_\gamma - \frac{1}{2} m_b) \tag{10}$$

which will be the leading term of the inclusive decay spectrum of $B \to \gamma X_s$ in $1/m_b$ expansion obviously. We write the decay width in the framework of the parton model

$$E_B \, d\Gamma(B \to \gamma X_s) = \int dx \, f(x) \, E_b \, d\Gamma(b \to s\gamma) \ . \tag{11}$$

The decay rate of the partonic subprocess is directly obtained from amplitude, $d\Gamma \propto |\mathcal{M}|^2 = G_F^2 e^2 |V_{tb} V_{ts}^*|^2 |c_7(m_b)|^2 m_b^6 / 4\pi^4$ since the amplitude is constant for 2–body decay. We have the photon energy spectrum

$$\frac{d\Gamma}{dE_\gamma} = \frac{2}{m_B} \Gamma_0 f(x_+) \left(1 + \mathcal{O}(\frac{1}{m_b})\right) \tag{12}$$

where $x_+ = 2E_\gamma / m_B$ and $\Gamma_0$ is defined in eq. (10). This spectrum is shown in Fig. 1. We note that the line shape is directly determined by the distribution function of $b$–quark inside $B$ meson as is observed in the Ref. [11,12].

In the framework of the HQET, Bigi et al. [11] derived the photon energy spectrum by

$$\frac{d\Gamma}{dE_\gamma} = \frac{2}{\bar{\Lambda}} \Gamma_0 F(x_B) \left(1 + \mathcal{O}(\frac{1}{m_b})\right) \tag{13}$$

where

$$x_B = \frac{2}{\bar{\Lambda}} \left(E_\gamma - \frac{1}{2} m_b\right) \ . \tag{14}$$

We find that the line shape is determined by the QCD distribution function $F(x_B)$, which shows scaling behaviors of QCD without perturbative QCD corrections. However the genuine function $F(x_B)$ cannot be determined completely by perturbative consideration. The HQET gives only some restrictions on this function, e.g. we can calculate moments of $F(x_B)$ in principle with the help of the HQET and OPE. Here we face the need of constructing realistic models with the help of informations from QCD.

We find that the result of parton model also shows the scaling properties and that the distribution function $f(x_+)$ is directly related the line shape function $F(x_B)$. By some variables changes, we obtain the relation

$$\frac{1}{m_B} f(x_+) = \frac{1}{\bar{\Lambda}} \left( \frac{\bar{\Lambda}}{m_B} \hat{f}(x_+ - x_0) \right) = \frac{1}{\bar{\Lambda}} F(x_B) \ . \tag{15}$$

The function $\hat{f}$ is the one obtained by the variable change $x_+ \to (x_+ - x_0) + x_0$ from $f(x_+)$. The functional form of $F(x_B)$ is obtained by the rescaling $(x_+ - x_0) = (\bar{\Lambda}/m_B) x_B$ from $\hat{f}(x_+ - x_0)$. Consequently the distribution function of QCD, $F(x_B)$, is the very distribution function of $b$–quark in our picture up to variables changes.

We show that in the inclusive semileptonic decay process, the differential decay rate of the parton model is also directly related to that of the HQET after the appropriate variable changes. The scaling behavior in the inclusive semileptonic decay process is derived in the HQET, see ref. [11]

$$\frac{d\Gamma}{dE_l dq^2 dq_0} = \frac{G_F^2 |V_{ub}|^2}{192\pi^3} \frac{2}{\bar{\Lambda}} F(\hat{x}_B) \frac{24(q_0 - E_l)(2m_b E_l - q^2)}{m_b - q_0} \tag{16}$$

where

$$\hat{x}_B = -\frac{m_b^2 + q^2 - 2m_b q_0}{2\bar{\Lambda}(m_b - q_0)} \tag{17}$$

Apart from kinematic factor, the triple differential decay rate is determined by one scaling variable $\hat{x}_B$. The scaling behaviors of QCD is violated by the perturbative and nonperturbative (higher twist) corrections.

The parton model approach as is presented in eq. (1) gives the triple differential decay spectrum

$$\frac{d\Gamma}{dE_l dq^2 dq_0} = \frac{G_F^2 |V_{ub}|^2}{4\pi^3} \frac{q_0 - E_l}{\sqrt{\mathbf{q}^2 + m_q^2}} \left( x_+ f(x_+)(2E_l - m_B x_-) + (x_+ \leftrightarrow x_-) \right) \tag{18}$$

where

$$x_{\pm} = \frac{q_0 \pm \sqrt{\mathbf{q}^2 + m_q^2}}{m_B} \quad .$$

Jin and Paschos [3] argued that the contribution of $f(x_-)$ is expected to be relatively small in the kinematic region where we are interested in and can be neglected by the kinematic analysis. This is in consequence of the fact the light–cone distribution function is sharply peaked around $x \sim m_b/m_B$. We let $m_q = 0$ hereafter. From the eq. (18), we obtain the coincident form with that of the HQET

$$\frac{d\Gamma}{dE_l dq^2 dq_0} \approx \frac{G_F^2 |V_{ub}|^2}{192\pi^3} \frac{2}{m_B} f(x_+) \frac{24(q_0 - E_l)(2m_b E_l - q^2)}{m_b - q_0} \quad . \tag{19}$$

As the case of $B \to \gamma X_s$, the line shape function $F(\hat{x}_B)$ is related to the distribution function $f(x_+)$ of our model in the same manner after appropriate variable changes. We see that the parton model shows the same form of scaling behavior as that of QCD. It would also be violated by the perturbative QCD corrections and nonperturbative effects which are mainly related to the behaviors far from the light–cone.

## IV. MOMENTS OF THE DISTRIBUTION FUNCTION

We have investigated that the first postulation of the parton model is consistent with that of the HQET. Now the validity of the model is attributed to the functional form of the distribution function $f(x)$. Here, we study the second postulate, the validity of Peterson's fragmentation function as a distribution function of $b$–quark. Using the HQET and the OPE, the moments of QCD distribution function

$$a_n = \int dx_B \, x_B^n F(x_B) \quad , \qquad n = 0, 1, ..., \tag{20}$$

are calculated from the expectation values of local operators between B mesons:

$$\frac{1}{2m_B} < B|S b \pi_{\mu_1}...\pi_{\mu_n} b - \text{traces}|B> = a_n \bar{\Lambda}^n (v_{\mu_1}...v_{\mu_1} - \text{traces}) \tag{21}$$

where $S$ denotes the symmetrization operator and $\pi_\mu = iD_\mu - m_b v_\mu$. $a_0 = 1$ from $b$–number conservation, and $a_1 = 0$ up to $\mathcal{O}(1/m_b^2)$ order indicates the lack of $1/m_b$ corrections. For the second moment of the spectrum, it is related to the average kinetic energy of $b$–quark inside the $B$ meson. We have

$$a_2 = \frac{1}{3\bar{\Lambda}^2} \frac{1}{2m_B} \langle B|\bar{b}\pi^2 b|B\rangle \equiv \frac{\mu_\pi^2}{3\bar{\Lambda}^2} \ . \tag{22}$$

With Peterson's function, the moments of the photon energy spectrum are numerically calculated. We have $a_0 = 1$ from normalization. $a_1$ is simply zero because we define the $b$–quark mass is the average value of the distribution function $f(x)$ times the meson mass $m_B$. In fact, our definition of $b$–quark mass aims at this feature of QCD. The second moment is derived as

$$a_2 = \int_{-\infty}^{1} dx_B \ x_B^2 F(x_B) = \left(\frac{m_B}{\bar{\Lambda}}\right)^2 \int_0^1 dy \ (y - \frac{m_b}{m_B})^2 f(y) \tag{23}$$

where $y = 2E_\gamma/m_B$. We obtain the numerical value $a_2 \simeq 0.757$ with the parameter $\epsilon = 0.004$ and the mass of $B$ meson $m_B = 5.3$ GeV, which well agrees with the result of the HQET and QCD sum rules, $a_2 \sim 0.5 - 1$. Thus one would conclude that the photon energy spectrum predicted from the parton model is in accord with the result of the HQET up to the order of $1/m_b^2$. This result explains the fact that the parton model present well–agreed spectrum with the experiment from ref. [2].

May we conclude that Peterson's form is appropriate as the genuine distribution function of $b$–quark inside the $B$ meson from this agreement? The answer is no. In our framework, the value of $b$–quark mass is defined as $m_b \simeq 4.5$ GeV and the effective mass $\bar{\Lambda} \simeq 0.8$ GeV. These values show some differences from those predicted from the HQET, $\bar{\Lambda} \simeq 0.4 - 0.6$ GeV. Moreover the average kinetic energy of $b$–quark is calculated in the parton model $\mu_\pi^2 = 1.43$ GeV$^2$, which is too much larger than the value estimated from QCD sum rules [13] to be $\mu_\pi^2 \sim 0.6$ GeV$^2$. The former discrepancy means that Peterson's function gives too small mean value and the latter that it gives too large variance. These results make the conclusion that Peterson's function does not give the correct description of $b$–quark distrbution inside $B$ meson comparing with that of real QCD. It gives allowable values of the second moment since above two effects are cancelled each other in calculation of $a_2$. We will discuss the improvement of Peterson's function as a distribution function in section 5.

## V. IMPROVEMENT OF PETERSON'S FUNCTION

We study Peterson's fragmentation function here. This function has been used as a description of fragmentation process of $b$- and $c$-quarks into hadrons containing one heavy flavour. The functional form is not purely ad hoc.. It relates the energy transfer to the fragmentation amplitudes. As the distribution function of $b$-quark inside $B$ meson, Peterson's function gives rather wrong values for some QCD calculations as shown in section 4. This function shows large asymmetric behavior with respect to the peak and it gives too small mean value which is too much deviated from the peak and too large width. We find that its functional form should be improved.

The improvement is performed by adding a following physical argument, which makes better the functional form of the Peterson's function as an QCD distribution function. As stated before, Peterson's function describes the fragmentation process by the energy transfer of the process. In the vewpoint of the parton model, the principal features of the amplitude for a fast moving heavy quark $Q$ fragmentation into a hadron $H$ and a light quark $q$ are determined by the value of the energy transfer $\Delta E = E_H + E_q - E_Q$ such that amplitude$(Q \to H + q) \propto \Delta E^{-1}$. On the other hand, we are considering the decay process in this paper. If we accept the assumption that the amplitude is determined by the energy transfer, the energy transfer should be $\Delta E = E_Q + E_q - E_H$ in decay preocess. This is the same as the fragmentation process when we let $m_B = m_b$ as Peterson et al. assumed [10]. We know that actually $m_B > m_b$ and we introduce a new parameter $\alpha_Q$ to express the mass difference such that $m_B^2 = m_b^2(1 + \alpha_Q)$. With this alteration, the energies about the transverse particle masses are given by

$$\Delta E = (m_b^2 + z^2 P^2)^{1/2} + (m_q^2 + (1-z)^2 P^2)^{1/2}$$

$$-(m_B^2 + P^2)^{1/2}$$
$$\propto 1 + \alpha_Q - \frac{1}{z} - \frac{\epsilon}{1-z} \tag{24}$$

where $z$ is a ratio of $b$-quark momentum to $B$ meson momentum. Then we obtain the function

$$f_Q^{\text{new}}(x) = \frac{N_Q}{x(1 + \alpha_Q - \frac{1}{x} - \frac{\epsilon_Q}{1-x})^2} \tag{25}$$

We show the functional form of this function in Fig. 2 compared with the original Peterson's form with the parameter $\epsilon_Q = 0.004$ and $\alpha_Q = 0.085$. As we expected, this new function gives larger mean value and smaller variance.

Using this improved function with the parameter $\epsilon_Q = 0.004$, $\alpha_Q = 0.085$, we obtain the numerical value for a few moments. Two moments $a_0$ and $a_1$ are not changed. The second moment $a_2 = 0.673$, which gives the proper line shape. We also obtain the values of the effective mass $\bar{\Lambda} = 0.537$ and $\mu_\pi^2 = 0.583$, which are consistent with those obtained by the HQET and QCD sum rules. Up to this order, the parton model with our improvement gives the consistent results with those of the genuine QCD distribution function estimated from QCD sum rules. We present the moments and the values of $\bar{\Lambda}$ and $\mu_\pi^2$ with varying $\alpha_Q$=0.02–0.1 in Table 1.

## VI. CONCLUDING REMARKS

We study the parton model with the heavy quark expansion through the inclusive radiative decay and the inclusive semileptonic decay of $B$ mesons. We properly define the mass of $b$–quark in the framework of this model and write the expansion of the distrbution function of $b$–quark in $1/m_b$. By relating the $b$–quark mass to the meson mass with this heavy quark expansion, we see that our definition of the $b$–quark mass satisfies the features presented by QCD.

In the parton model, the distribution function of $b$–quark, which describe the motion of the quark inside the $B$ meson, plays a major role. It has been shown that the distribution function $f(x)$ is directly related to the QCD distribution function determining the line shape of the decay spectra. Here we use Peterson's fragmentation function as distribution function of $b$–quark following Ref. [1,2]. However the original Peterson's function gave rather wrong values for the average kinetic energy of $b$–quark inside $B$ meson and the effective mass of the light degrees of freedom while it gave allowable numerical values for the second moment $a_2$. Thus improvement of the function is required. We observe that Peterson's function assumes that $m_b = m_B$. By parametrizing the mass difference of $b$–quark and $B$ meson, we rederived the distribution function for decay process and obtained the refined form eq. (24). We showed that this function satisfies the restriction of QCD and conclude that the model is successfully improved.

In Fig. 3, we compare the decay spectra calculated by Peterson's function with those by our refined function. We find that the lepton energy spectrum of $B \to e\bar{\nu}X_u$ decay does not show mush change in spite of the drastic change of the distribution function. This resulted in the fact that Jin et al. could obtain well agreed spectrum with experiments by using Peterson's function as the distribution function in ref. [2].

## ACKNOWLEDGMENTS

K. Y. L particularly thanks to Y. G. Kim for helpful discussions and C. S. Kim for valuable comments. This work is supported in part by the Korean Science and Engineering Foundation.

# REFERENCES

[1] A. Bareiss and E. A. Paschos, Nucl. Phys. **B 327** (1989) 353.

[2] C. H. Jin, W. F. Palmer and E. A. Paschos, Phys. Lett. **B 329** (1994) 364; C. H. Jin, W. F. Palmer and E. A. Paschos, Preprint DO - TH 93/21 (1993) (unpublished): C. H. Jin, W. F. Palmer and E. A. Paschos, Preprint DO - TH 94/12 (1994).

[3] C. H. Jin and E. A. Paschos, Preprint DO - TH 95/07 (1995).

[4] J. Chay, H. M. Georgi and B. Grinstein, Phys. Lett. **B 247** (1990) 399; I. Bigi, M. Shifman, N. G. Uraltsev and A. I. Vainshtein, Phys. Rev. Lett. **71** (1993) 496.

[5] A. V. Manohar and M. B. Wise, Phys. Rev. **D 49** (1994) 1310.

[6] B. Blok, L. Koyrakh, M Shifman and A. I. Vainshtein, Phys. Rev. **D 49** (1994) 3356.

[7] G. Altarelli, N. Cabbibo, G. Corbo, L. Maiani and G. Martinelli, Nucl. Phys. **B 208** (1982) 365.

[8] G. Baillie, Phys. Lett. **B 324** (1994) 446; C. Csáki and L. Randall, Phys. Lett. **B 324** (1994) 451.

[9] I. I. Bigi, M. A. Shifman, N. G. Uraltsev and A. L. Vainshtein, Phys. Lett. **B 328** (1994) 431.

[10] C. Peterson, D. Schlatter, I. Schmitt and P. M. Zerwas, Phys. Rev. **D 27** (1983) 105.

[11] I. I. Bigi, M. A. Shifman, N. G. Uraltsev and A. L. Vainshtein, Int. J. Mod. Phys. **A 9** (1994) 2467.

[12] M. Neubert, Phys. Rev. **D 49** (1994) 3392; 4623.

[13] P. Ball and V. M. Braun, Phys. Rev. **D 49** (1994) 2472.